\documentclass[11pt,a4paper]{article}
\usepackage{amsmath,amssymb}
\usepackage{epsfig,graphicx}

\begin{document}

\title{\bf Neutrino dispersion in  magnetized plasma }
\author{N.~V.~Mikheev\footnote{{\bf e-mail}: mikheev@uniyar.ac.ru},
E.~N.~Narynskaya\footnote{{\bf e-mail}: elenan@uniyar.ac.ru}
\\
\small{\em Yaroslavl State University } \\
\small{\em  150000 Russian Federation, Yaroslavl, Sovietskaya 14.}
}
\date{}
\maketitle

\begin{abstract}
The neutrino dispersion in the charge symmetric magnetized plasma is investigated. We have studied 
the plasma contribution into the additional energy of neutrino and obtained the simple expression for it.
We consider in detail
the neutrino self-energy
 under physical conditions of weak field, moderate field and strong field limits. It is shown that our result for neutrino dispersion in moderate magnetic field differ substantially from the previous one in the  literature.
\end{abstract}

\def\beq{\begin{equation}}
\def\eeq{\end{equation}}
\def\bd{\begin{displaymath}}
\def\ed{\end{displaymath}}

\section{Introduction.}

\indent\indent 
The investigations of neutrino physics in an active medium are the subject of great interest today. The one of the topical problem in current researches is the influence of an external magnetic field and plasma on the neutrino dispersion relation.

It this paper we analyse  the neutrino dispersion properties in an active medium consisting  of magnetic field and plasma. The investigations of such type are based on calculation of neutrino self-energy operator $\Sigma (p)$. This operator can be defined in term of the  invariant amplitude of the neutrino transition $\nu \to \nu$, by the relation:
\beq
M(\nu \to \nu) = - \, \bar\nu(p) \, \Sigma(p)\,\nu(p),
\label{eq:def}
\eeq
where $p^\mu$ is the neutrino four-momentum. 

Using the expression (\ref{eq:def}), one obtains for the additional neutrino energy in magnetized plasma~\footnote
{ We use natural units in which
$c=\hbar=1$.}:
\beq
\Delta E = \frac{1}{4E}\, Sp \left \{ ((p\gamma) + m_\nu)\,(1 - (s\gamma)\,\gamma_5)\,\Sigma (p) \right \},
\label{eq:2}
\eeq
where $E$ is the neutrino energy in vacuum, $m_\mu$ is the neutrino mass, $s^\mu$ is the neutrino spin four-vector,
$\gamma^\alpha$ are the Dirac matrices in the standard presentation. The Lorentz indexes 
 of four-vectors and tensors within parenthesis are contracted consecutively, for example, $(p\gamma)=p^\mu \gamma_\mu$.

There are several parametrization for general structure of the operator $\Sigma(p)$. In the presence 
of homogeneous magnetized medium the operator $\Sigma(p)$ contains three independent structures. In this case
it is convenient to express $\Sigma(p)$  as
\beq
\Sigma(p) = \left \{ a \,(p\gamma) + b \,(u\gamma) + c \,(p\tilde\varphi \gamma)\right \}\, L.
\label{eq:3}
\eeq

Here $a,b,c$ are the numerical coefficients, $u^\mu$ is the four-vector of medium velocity,
$\tilde\varphi_{\alpha\beta}=\tilde F_{\alpha\beta}/B$ is the dimensionless dual tensor of the magnetic field,
$B$ is the absolute value of the magnetic field strength,
$L=(1-\gamma_5)/2$ is the left-handed projection operator. 

On this parametrization the coefficients $a,b,c$ have a simple physical interpretation. Really, performing calculation of the trace in eq.(\ref{eq:3}) with $\Sigma(p)$ in the form (\ref{eq:2}), one can obtained for
 the neutrino self-energy
\beq
\Delta E = b \,\, \frac{(1 - (\vec v \vec \xi))}{2 } - c \, \,\frac{m_\nu}{2E}\,(p \tilde\varphi s).
\label{eq:4}
\eeq
where $\vec v$ is the neutrino velocity, $\vec \xi$ is the twice vector of average  spin of neutrino.

It is seen that the additional energy in magnetized plasma for massless left-handed
neutrino in fact depends on the parameter \,$b$ \,only,
\bd
\Delta E = b.
\ed
In the case of massive neutrino the second term in (\ref{eq:4}) corresponds to the additional 
energy caused by  a neutrino magnetic moment $\mu_\nu$, 
\bd
\Delta E_{\mu_\nu} = - \frac{\mu_\nu \, B}{E} \,(p \tilde\varphi s), \qquad  \mu_\nu = \frac{c \,m_\nu}{2 B}.
\ed
So, one can speak that parameter \,$c$ \, determines the additional  neutrino magnetic moment
 in the magnetized plasma.

The modification of the neutrino dispersion relation in a magnetized plasma have a long history, see for example~\cite{Raffelt}-\cite{Mikheev}. In particular, the dispersion relation of neutrino~\footnote{we consider
the electron neutrino} in charge symmetric plasma under physical conditions
\vspace{2mm}
\beq
m_W^2 \gg T^2, \, eB \gg m_e^2, \qquad eB \le T^2
\label{eq:cond}
\eeq
\vspace{2mm}
 is well known
 \vspace{2mm}
 \beq
 \frac{\Delta E}{|\vec p|} = \frac{\sqrt{2}\,G_F}{3} \left [ - \frac{7 \, \pi^2 \,T^4}{15} \left ( 
\frac{1}{m^2_Z} + \frac{2}{m_W^2}\right ) + \frac{T^2 \,eB}{m_W^2}\, \cos\phi + \frac{(eB)^2}{2 \,\pi^2 \,m^2_W}\,
\ln\left ( \frac{T^2}{m^2_e}\right) \, \sin^2\phi\, \right ].
\label{eq:5}
\eeq

 \vspace{4mm}
 
 Here $\vec p$ is the neutrino momentum, $\phi$ is the angle between magnetic field direction and vector $\vec p$,
 $T$ is the plasma temperature. 
 
 The first term in (\ref{eq:5}) is the pure plasma contribution~\cite{Raffelt}, while the second~\cite{Elmfors} and third~\cite{Erdas} terms in brackets are caused by the common influence of the plasma and magnetic field. As one can see the term of the second order of the field  contains the infrared divergence in the massless electron limit. However, there is a good reason to think that this large logarithmic factor $\ln (T^2/m^2_e)$ can not arise under physical conditions (\ref{eq:cond}),
 when the electron mass is the smallest parameter of the task. Actually, under conditions (\ref{eq:cond}) 
 the contribution into neutrino energy is determined by electrons and positrons on exited Landau levels with energy
 $\omega_n = \sqrt{k_3^2 + 2eBn + m^2_e}$. One can see  that in this case the  electron mass squared 
 can be neglected in comparison to magnetic field. By this means, the availability of the
 logarithmic factor $\ln (T^2/m^2_e)$ in the (\ref{eq:5}) has come into question, so
 independent  investigations of the neutrino dispersion in magnetized medium it is required. 
  
  Further we study the additional neutrino energy for massless neutrino in the
   charge symmetric plasma with the presence of the
    magnetic field of  arbitrary strength. As a special  case we  consider the weak field, moderate field and strong field limits in more details.

\section{The plasma contribution into neutrino self-energy operator $\Sigma(p)$.}
\indent\indent

In the case of the charge symmetric plasma 
 the contribution into amplitude of neutrino transition $\nu \to \nu$ in magnetized plasma  derives
 from  processes, shown in fig.1.  It should be noted that in the local limit the 
neutrino self-energy operator $\Sigma(p)$ is zero~\cite{Raffelt}. Thus,  there is a need to take into account the 
momentum dependence  of W,Z - boson propagators. 
 The process of neutrino transition via
 Z boson does not sensitive to influence of external magnetic field.
 Thus, the contribution into operator $\Sigma (p) $ from this process can 
 be extracted from paper~\cite{Raffelt}. 
 
 The amplitude of neutrino forward scattering on all plasma electrons and positrons can be 
 immediately obtained from the Lagrangian of $\nu e$ interaction 
 with W-boson
 \beq
 L = \frac{g}{2\sqrt{2}} \left ( \bar e \,\gamma_\alpha\,(1 - \gamma_5)\,\nu \, \right )\,W_\alpha +
       \frac{g}{2\sqrt{2}} \left ( \bar \nu_e \,\gamma_\alpha\,(1 - \gamma_5)\, e \,\right)\, W^*_\alpha.
       \label{lag}
       \eeq

\begin{figure}[t]
\centerline{\includegraphics{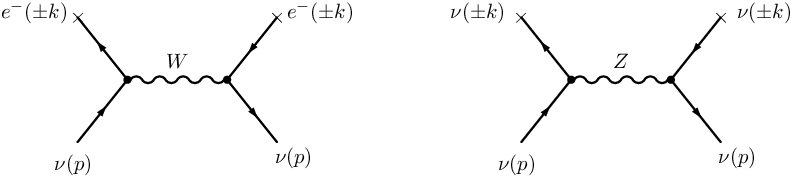}}
\caption{Feynman diagrams for the plasma contribution into the neutrino self-energy operator.}
\vspace{8mm}
\label{fig}
\end{figure}

Omitting the details of calculations, the general expression for the neutrino self-energy  caused by the neutrino forward
scattering on plasma electrons can be written as
\begin{eqnarray}
\frac{\Delta E}{|\vec p|} & = & -\frac{2\sqrt{2}\, G_F\,eB}{\pi^2\, m^2_W}\,\int\limits^{+\infty}_{-\infty}
\frac{dk_3\, f(\omega_n)}{\omega_n} \times \label{eq:total} \\
& \times & \left ( \sum^{\qquad '}_{n=0} \, (\omega^2_n + eBn + \cos^2\phi\,(k^2_3 - eBn)) - 
\frac{\delta_{n0}}{2} \, \cos\phi \, (k^2_3 + \omega_n^2) \right ). \nonumber
\end{eqnarray}

Here $\omega_n = \sqrt{k_3^2 + 2eBn + m^2_e}$ and $k_3$ are the electron energy and
 z-component~\footnote{the magnetic field is directed along the z-axis } of electron
momentum  correspondingly,
 $n$ is the Landau level number, $\phi$ s the angle between magnetic field direction and neutrino momentum $\vec p$,
 $f(\omega_n)$  is the electron distribution function, 
 $f(\omega_n)=[exp(\,\omega_n/T\,)-1]^{-1}$,  the sum is defined as
 \bd
 \sum^{\qquad '}_{n=0} F(n) = \frac{1}{2} \,\, F(n=0) + \sum^{\infty}_{n=1} F(n).
 \ed

The integral and sum in eq. (\ref{eq:total}) can be calculated under some physical conditions:

\begin{itemize}

\item the limit of weak magnetic field, when the magnetic field strength is the smallest physical parameter
\beq
T^2 \gg m^2_e \gg eB.
\label{eq:cond1}
\eeq
 The result of  calculations for the plasma contribution into neutrino self-energy in this limit
  is
\begin{eqnarray}
\frac{\Delta E}{|\vec p\,|} & = & \frac{\sqrt{2}\,G_F}{3 m^2_W} \left [ - \frac{7 \, \pi^2 \,T^4}{15} \left ( 
2 + \frac{m_W^2}{m^2_Z}\right ) +  T^2 \,eB\, \cos\phi +  \right .\label{eq:weak} \\
& \qquad & \left . \qquad\qquad\qquad\qquad\qquad\qquad\qquad + \,\frac{(eB)^2}{2 \,\pi^2}\,
\left \{\sin^2\phi \left (  \ln\left (\frac{T^2}{m^2_e}  \right) + 0,635 \right ) - 1 \right \} \, \right ].
\nonumber
\end{eqnarray}

  As one can see the expression (\ref{eq:weak}) contains the logarithmic factor with $m_e$, but under conditions 
  considered (\ref{eq:cond1}) the electron mass is not the smallest parameter of task. So the electron mass
  can not be tend to zero at nonzero magnetic field B.

 \item the moderate magnetic field, when the magnetic field being relatively weak in comparison with 
 plasma temperature, is simultaneously strong enough on the scale of electron mass squared
 \beq
 T^2 \gg eB \gg m^2_e.
 \label{eq:cond2}
 \eeq
  Under conditions (\ref{eq:cond2}) plasma electrons and positrons occupy highest Landau levels. In this case for neutrino energy one can obtained
\begin{eqnarray}
\frac{\Delta E}{|\vec p\,|} & = & \frac{\sqrt{2}\,G_F}{3 m^2_W} \left [ - \frac{7 \, \pi^2 \,T^4}{15} \left ( 
2 + \frac{m_W^2}{m^2_Z}\right ) +  T^2 \,eB\, \cos\phi + \right .\label{eq:moderate} \\
& \qquad & \left . \qquad\qquad\qquad\qquad\qquad\qquad\qquad + \,\frac{(eB)^2}{2 \,\pi^2}\,
\left \{\sin^2\phi \left (  \ln\left (\frac{T^2}{eB}  \right) + 2,93 \right ) - 1 \right \} \, \right ].
\nonumber
\end{eqnarray}

 Our calculations show, that under conditions (\ref{eq:cond2})
  the additional neutrino energy does not contain the infrared divergence in the limit $m_e \to 0$ in contrast~\cite{Erdas}.
 
  \item the strong magnetic field limit, when from two components of active medium the field component dominates
 \beq
 eB \gg T^2 \gg m^2_e.
 \label{eq:cond3}
 \eeq

 Under conditions (\ref{eq:cond2}) the most part of plasma electrons and positrons occupy the ground Landau level.
 The result  for the additional neutrino energy in this limit has the form: 
\begin{eqnarray}
\frac{\Delta E}{|\vec p\,|} & = & - \frac{\sqrt{2}\,G_F}{3 m^2_W} \left [\frac{7 \, \pi^2 \,T^4\,m^2_W}{15 \, m^2_Z}
 +  \frac{T^2 \,eB}{2}\, (1 - \cos\phi)^2 + \label{eq:strong} \right . \\
& \qquad & \left . \qquad\qquad\qquad\qquad\qquad+ \,
3\, (eB)^2 \, \left ( \frac{2}{\pi}\right )^{3/2}\, \left ( \frac{T^2}{2eB}\right )^{1/4}\,
( 3 - \cos^2\phi) \, e^{-\sqrt{2eB/T}}\, \right ]. \nonumber
\end{eqnarray}

 Here the second term corresponds to the contributions of ground Landau level, while the third term is caused by the 
 first Landau level.

\end{itemize}
\section{Conclusion.}
\indent\indent 
We have studied the neutrino dispersion in the charge symmetric plasma with the presence of a constant
magnetic field. The most general expression in simple analytic form 
for the plasma contribution into the neutrino self-energy was obtained. In particular, 
we have considered the physical conditions, corresponding to the weak field and moderate field,
when plasma electrons and positrons occupy the excited Landau levels. 
The strong magnetic field limit, when plasma electrons 
and positrons mainly occupy the lowest Landau level, is investigated  also.

It is shown that additional neutrino energy in the limit of  moderate field,
$T^2 \gg eB \gg m^2_e$, does not coincide with previous result in paper ~\cite{Erdas} under the
same physical conditions and does not contain the infrared divergence in the limit $m_e \to 0$ in contrast to~\cite{Erdas}.

\vspace{5mm}

The work was supported in part by the Russian Foundation for Basic Research
under the Grant No.~07-02-00285-a, and by the Council on Grants by the
President
of the Russian Federation for the Support of Young Russian Scientists
and Leading Scientific Schools of Russian Federation under the Grant
No.~NSh-497.2008.2.

\vspace{5mm}



\begin{thebibliography}{99}


\bibitem{Raffelt}
D.Notzold, G.Raffelt,
   Nucl.Phys.B   \textbf{307}, 924, (1988).
  
\bibitem{Elmfors}
P.~Elmfors, D.~Grasso and G.Raffelt,
   Nucl.Phys.B   \textbf{479}, 3, (1996).


\bibitem{Erdas}
A.~Erdas, G.W.Kim, T.H.Lee, Phys. Rev. D \textbf{58}, 085016, (1998).

\bibitem{Elizalde}
E.~Elizalde, E.J.Ferrer, V. de la Incera, Phys. Rev. D \textbf{70}, 043012, (2004).

\bibitem{Mikheev}
A.V.~Kuznetsov, N.V.Mikheev, G.G.Raffelt,
L.A.Vassilevskaya, Phys. Rev. D \textbf{73}, 023001, (2006).

\end{thebibliography}
\end{document}